\documentclass[preprint,showpacs,byrevtex,groupedaddress,pra,10pt,twocolumn]{revtex4}
\usepackage{amsfonts}
\usepackage{amsmath}
\usepackage{amssymb}
\usepackage{graphicx}
\usepackage{dcolumn}
\usepackage{bm}

\begin{document}

\preprint{}
\title[Density-functional calculations of inner-shell excitation]{%
Spin-dependent localized Hartree-Fock density-functional approach for the
accurate treatment of inner-shell excitation of close-shell atoms}
\author{Zhongyuan Zhou$^{1,2}$}
\author{Shih-I Chu$^{1}$}
\affiliation{$^{1}$Department of Chemistry, University of Kansas, Lawrence, KS 66045\\
$^{2}$Department of Physics and Astronomy, University of Kansas, Lawrence,
KS 66045}
\keywords{one two three}
\pacs{31.15.Ew, 32.80.Wr, 32.80.Rm}

\begin{abstract}
We present a spin-dependent localized Hartree-Fock (SLHF) density-functional
approach for the treatment of the inner-shell excited-state calculation of
atomic systems. In this approach, the electron spin-orbitals in an
electronic configuration are obtained first by solving Kohn-Sham (KS)
equation with SLHF exchange potential. Then a single-Slater-determinant
energy of the electronic configuration is calculated by using these electron
spin-orbitals. Finally, a multiplet energy of an inner-shell excited state
is evaluated from the single-Slater-determinant energies of the electronic
configurations involved in terms of Slater's diagonal sum rule. This
procedure has been used to calculate the total and excitation energies of
inner-shell excited states of close-shell atomic systems: Be, B$^{+}$, Ne,
and Mg. The correlation effect is taken into account by incorporating the
correlation potentials and energy functionals of Perdew and Wang's (PW) or
Lee, Yang, and Parr's (LYP) into calculation. The calculated results with
the PW and LYP energy functionals are in overall good agreement with each
other and also with available experimental and other \textit{ab initio}
theoretical data. In addition, we present some new results for highly
excited inner-shell states.
\end{abstract}

\received{June 17, 2006}
\maketitle

\section{Introduction}

Density functional theory (DFT) \cite{Hohenberg64,Kohn65} has been widely
applied to many areas in theoretical physics and chemistry as a powerful 
\textit{ab initio} approach for the calculation of ground-state properties
of many-electron systems due to its computational simplicity and efficiency 
\cite{Parr89,Dreizler90}. The basic equation of DFT is Kohn-Sham (KS)
equation \cite{Kohn65} and the key part in KS equation is
exchange-correlation (XC) potential \cite{a16}.

DFT with a traditional XC potential obtained from uniform electron gas, such
as local density approximation (LDA) \cite{Parr89,Dreizler90} and
generalized gradient approximation (GGA) \cite{Becke88,Perdew86,a38,a21}, is
a ground-state approach. Because of incomplete cancellation of spurious
self-interactions in the conventional DFT using LDA or GGA \cite%
{Parr89,Perdew86,a38,a21} and the inherent degeneracy (due to the use of
spin and angular-momentum independent local potentials), the differences of
the KS energy eigenvalues of unoccupied and occupied orbitals are not
rigorously defined as excitation energies. However, the KS energy
eigenvalues can serve as good zeroth-order excited-state energies provided
they are obtained by solving KS equation with a high-quality XC potential 
\cite{a15}. A number of theoretical methods have been developed by adopting
this point of view \cite{Singh99}. In particular, density work-functional
approach (WF) \cite{a11,a13,a14,a39}, open-shell localized Hartree-Fock
(LHF) density-functional approach \cite%
{Sala2003,Sala2003-1,Vitale05,Gorling05}, and multireference LHF
density-functional approach \cite{Hupp2003,Hupp2003-1}, etc., have been
successfully used to calculate excited-state properties of atomic and
molecular systems.

Recently, an exchange (X)-only LHF density-functional theory has been
proposed and successfully applied to ground-state calculations of atomic and
molecular systems \cite{a15}. In this X-only DFT, the exchange potential in
the KS equation is a LHF exchange potential derived under the assumption
that X-only KS determinant is equal to the Hartree-Fock (HF) determinant. We
have recently extended this approach to excited states of atomic and
molecular systems by assuming that the X-only KS determinant is also equal
to the HF determinant for excited states \cite{zhou2005}. Based on this
postulate we have developed a spin-dependent localized Hartree-Fock (SLHF)
density-functional approach for excited-state calculation of atomic and
molecular systems \cite{zhou2005}. In this approach, the exchange potential
in the KS equation is an exact nonvariational SLHF exchange potential
constructed for both the ground and excited states. The SLHF potential is an
analogue of the LHF potential. It is self-interaction free and exhibits the
correct long-range behavior. Further, the SLHF potential requires the use of
only the occupied orbitals and is dependent of the orbital symmetry of the
state. This approach associating with Slater's diagonal sum rule \cite%
{Slater60} has been successfully used to calculate singly, doubly, and
triply excited states of valence electrons of He- and Li-like ions \cite%
{zhou2005} with accurate results.

In this paper, we extend the SLHF density-functional approach to inner-shell
excited states of atomic systems. We compute the total and excitation
energies of inner-shell excited states of close-shell atomic systems: Be, B$%
^{+}$, Ne and Mg. In the calculation, the correlation potentials and energy
functionals proposed by Perdew and Wang (PW) \cite{a21} and\ by Lee, Yang,
and Parr (LYP) \cite{a38} are used to take into account the electron
correlation effect. We will show that the calculated results are in overall
good agreement with available theoretical and experimental data,
demonstrating that the SLHF density-functional approach can provide a simple
and computationally efficient approach for the accurate calculation of
inner-shell excited states of close-shell atomic systems within DFT.
Finally, we also present some new results for the highly excited inner-shell
states for the first time.

\section{Theoretical Method}

The SLHF density-functional approach has been discussed in Ref. \cite%
{zhou2005} in detail and is outlined in this section for convenience.

In spin-dependent density-functional approach, a spin-orbital $\varphi
_{i\sigma }\left( \mathbf{r}\right) $ of the $i$th electron with spin $%
\sigma $ ($\sigma =$ $\alpha $ and $\beta $ for spin-up and spin-down,
respectively) and its orbital energy $\varepsilon _{i\sigma }$ are
determined by the KS equation%
\begin{equation}
H_{\sigma }(\mathbf{r})\varphi _{i\sigma }\left( \mathbf{r}\right)
=\varepsilon _{i\sigma }\varphi _{i\sigma }\left( \mathbf{r}\right) ,
\label{e11}
\end{equation}%
where,%
\begin{equation}
H_{\sigma }(\mathbf{r})=-\frac{1}{2}\nabla ^{2}+V_{\sigma }^{\text{eff}%
}\left( \mathbf{r}\right) ,  \label{e11-1}
\end{equation}%
is the KS Hamiltonian and%
\begin{equation}
V_{\sigma }^{\text{eff}}\left( \mathbf{r}\right) =V_{ext}\left( \mathbf{r}%
\right) +V_{H}\left( \mathbf{r}\right) +V_{xc\sigma }\left( \mathbf{r}%
\right) ,  \label{e11-2}
\end{equation}%
is the local effective potential. In Eq. (\ref{e11-2}), $V_{ext}\left( 
\mathbf{r}\right) $ is the external potential, $V_{H}\left( \mathbf{r}%
\right) $ is Hartree potential (classical Coulomb electrostatic potential
between electrons), and $V_{xc\sigma }\left( \mathbf{r}\right) $ is the XC
potential.

For a given atomic system, the external potential $V_{ext}\left( \mathbf{r}%
\right) $ is known exactly. The Hartree potential $V_{H}\left( \mathbf{r}%
\right) $ is given by%
\begin{equation}
V_{H}\left( \mathbf{r}\right) =\int \frac{\rho \left( \mathbf{r}^{\prime
}\right) }{\left\vert \mathbf{r}-\mathbf{r}^{\prime }\right\vert }d\mathbf{r}%
^{\prime },  \label{e13}
\end{equation}%
where, $\rho \left( \mathbf{r}\right) =\rho _{\alpha }\left( \mathbf{r}%
\right) +\rho _{\beta }\left( \mathbf{r}\right) $ is the total electron
density and $\rho _{\sigma }\left( \mathbf{r}\right) $ (for $\sigma =\alpha $
and $\beta $) is the spin-dependent electron density defined by 
\begin{equation}
\rho _{\sigma }\left( \mathbf{r}\right) =\sum_{i=1}^{N_{\sigma }}w_{i\sigma
}\left\vert \varphi _{i\sigma }\left( \mathbf{r}\right) \right\vert ^{2}.
\label{e10}
\end{equation}%
Here $N_{\sigma }$ is the number of electrons with spin $\sigma $ and $%
w_{i\sigma }$ is the occupied number of electrons in the spin-orbital $%
\varphi _{i\sigma }\left( \mathbf{r}\right) $.

The XC potential can be decomposed into the exchange potential $V_{x\sigma
}\left( \mathbf{r}\right) $ and the correlation potential $V_{c\sigma
}\left( \mathbf{r}\right) $. In the SLHF density-functional approach, the
exchange potential is a SLHF exchange potential $V_{x\sigma }^{\text{SLHF}}(%
\mathbf{r})$. It is given by%
\begin{equation}
V_{x\sigma }^{\text{SLHF}}(\mathbf{r})=V_{x\sigma }^{\text{S}}(\mathbf{r}%
)+V_{x\sigma }^{\text{C}}(\mathbf{r}),  \label{e109}
\end{equation}%
where,%
\begin{equation}
V_{x\sigma }^{\text{S}}(\mathbf{r})=-\frac{1}{\rho _{\sigma }(\mathbf{r})}%
\sum_{i,j=1}^{N_{\sigma }}\gamma _{ij}^{\sigma }\left( \mathbf{r}\right)
\int \frac{\gamma _{ij}^{\sigma }\left( \mathbf{r}^{\prime }\right) }{%
\left\vert \mathbf{r}-\mathbf{r}^{\prime }\right\vert }d\mathbf{r}^{\prime },
\label{e110}
\end{equation}%
is the Slater potential \cite{Slater60} and%
\begin{equation}
V_{x\sigma }^{\text{C}}(\mathbf{r})=\frac{1}{\rho _{\sigma }(\mathbf{r})}%
\sum_{i,j=1}^{N_{\sigma }}\gamma _{ij}^{\sigma }\left( \mathbf{r}\right)
Q_{ij}^{\sigma },  \label{e111}
\end{equation}%
is a correction to Slater potential. In Eqs. (\ref{e110}) and (\ref{e111}) $%
\gamma _{ij}^{\sigma }\left( \mathbf{r}\right) $ and $Q_{ij}^{\sigma }$ are
defined by%
\begin{equation}
\gamma _{ij}^{\sigma }\left( \mathbf{r}\right) =\varphi _{i\sigma }(\mathbf{r%
})\varphi _{j\sigma }(\mathbf{r}),  \label{e111-1}
\end{equation}%
and%
\begin{equation}
Q_{ij}^{\sigma }=\left\langle \varphi _{j\sigma }\left\vert V_{x\sigma }^{%
\text{SLHF}}-V_{x\sigma }^{\text{NL}}\right\vert \varphi _{i\sigma
}\right\rangle ,  \label{e111-2}
\end{equation}%
where, $V_{x\sigma }^{\text{NL}}$ is a nonlocal exchange operator of the
form of HF exchange potential but constructed from KS spin-orbitals.

The SLHF exchange potential determined by Eqs. (\ref{e109})--(\ref{e111-2})
has two arbitrary additive constants. The physical orbitals can only be
obtained by the use of appropriate constants in the exchange potential \cite%
{a15}. To settle down the constants so as to pick up the physical orbitals,
it is required that the highest-occupied-orbital $N_{\sigma }$ of each spin $%
\sigma $ does not contribute to the correction term $V_{x\sigma }^{\text{C}}(%
\mathbf{r})$. In this case, the correction term $V_{x\sigma }^{\text{C}}(%
\mathbf{r})$ decays exponentially, the SLHF exchange potential behaves
asymptotically as Slater potential and thus approaches to $-1/r$ at long
range \cite{a15}.

In atomic systems, an electron spin-orbital is characterized by three
quantum numbers $n,$ $l,$ and $\sigma $, where $n$ and $l$ are the principal
quantum number and orbital angular momentum quantum number of the electron,
respectively. In the spherical coordinates, the spin-orbital $\varphi
_{i\sigma }\left( \mathbf{r}\right) $ of an electron with quantum numbers $%
n, $ $l,$ and $\sigma $ can be expressed by%
\begin{equation}
\varphi _{i\sigma }\left( \mathbf{r}\right) =\frac{R_{nl\sigma }(r)}{r}%
Y_{lm}(\theta ,\phi ),  \label{e204}
\end{equation}%
where, $R_{nl\sigma }(r)$ is the radial spin-orbital, $Y_{lm}(\theta ,\phi )$
is the spherical harmonic, $m$ is the azimuthal quantum number, and $i$ is a
set of quantum numbers apart from spin $\sigma $ of the spin-orbital. The
radial spin-orbital $R_{nl\sigma }(r)$ is governed by radial KS equation,%
\begin{equation}
\left[ -\frac{1}{2}\frac{d^{2}}{dr^{2}}+\frac{l(l+1)}{2r^{2}}+v_{\sigma }^{%
\text{eff}}(r)\right] R_{nl\sigma }=\varepsilon _{nl\sigma }R_{nl\sigma },
\label{e205}
\end{equation}%
where $v_{\sigma }^{\text{eff}}(r)$ is the radial effective potential given
by 
\begin{equation}
v_{\sigma }^{\text{eff}}(r)=v_{ext}\left( r\right) +v_{H}\left( r\right)
+v_{x\sigma }^{\text{SLHF}}\left( r\right) +v_{c\sigma }\left( r\right) .
\label{e206-1}
\end{equation}%
In Eq. (\ref{e206-1}), $v_{ext}\left( r\right) $, $v_{H}\left( r\right) $, $%
v_{x\sigma }^{\text{SLHF}}\left( r\right) $, and $v_{c\sigma }\left(
r\right) $ are the radial external potential, radial Hartree potential,
radial SLHF exchange potential, and radial correlation potential,
respectively.

For an atomic system with a nuclear charge $Z$, the external potential is
the Coulomb potential between electron and nucleus%
\begin{equation}
v_{ext}\left( r\right) =-\frac{Z}{r}.  \label{e207}
\end{equation}

In central-field approach, the radial Hartree potential is calculated from%
\begin{equation}
v_{H}\left( r\right) =4\pi \int \frac{1}{r_{>}}\rho (r^{\prime })r^{\prime
2}dr^{\prime },  \label{e210}
\end{equation}%
where, $r_{>}$ is the larger of $r$ and $r^{\prime }$, $\rho (r)=\rho
_{\alpha }(r)+\rho _{\beta }(r)$ is the spherically averaged total electron
density, and $\rho _{\sigma }(r)$ $\left( \sigma =\alpha \text{ or }\beta
\right) $ is the spherically averaged spin-dependent electron density given
by%
\begin{equation}
\rho _{\sigma }(r)=\frac{1}{4\pi }\int \rho _{\sigma }(\mathbf{r})d\Omega =%
\frac{1}{4\pi }\sum_{nl}^{\nu _{\sigma }}w_{nl\sigma }\left[ \frac{%
R_{nl\sigma }}{r}\right] ^{2}.  \label{e204-1}
\end{equation}%
Here the symbol $\nu _{\sigma }$ stands for a set of quantum numbers for
summation and the sum is performed over all the occupied spin-orbitals with
spin $\sigma $. This expression is accurate for spherically symmetric
(close-shell) states, but it is only an approximation for non-spherically
symmetric (open-shell) states. It may induce an error when it is used to
evaluate the energy of a non-spherically symmetric state. However, the error
is negligible compared to the order of calculated multiplet splitting \cite%
{Singh99}.

The radial SLHF exchange potential is given by%
\begin{equation}
v_{x\sigma }^{\text{SLHF}}\left( r\right) =v_{x\sigma }^{\text{S}}\left(
r\right) +v_{x\sigma }^{\text{C}}\left( r\right) ,  \label{e211}
\end{equation}%
where,%
\begin{equation}
v_{x\sigma }^{\text{S}}\left( r\right) =-\frac{1}{4\pi \rho _{\sigma }(r)}%
\sum_{nlm}^{\nu _{\sigma }}\sum_{n^{\prime }l^{\prime }m^{\prime }}^{\nu
_{\sigma }}s_{nlm,n^{\prime }l^{\prime }m^{\prime }}^{\sigma }(r),
\label{e212-1}
\end{equation}%
is the radial Slater potential and%
\begin{equation}
v_{x\sigma }^{\text{C}}\left( r\right) =\frac{1}{4\pi \rho _{\sigma }(r)}%
\sum_{nlm}^{\nu _{\sigma }}\sum_{n^{\prime }l^{\prime }m^{\prime }}^{\nu
_{\sigma }}c_{nlm,n^{\prime }l^{\prime }m^{\prime }}^{\sigma }(r).
\label{e213}
\end{equation}%
is a correction to the radial Slater potential. The matrix elements $%
s_{nlm,n^{\prime }l^{\prime }m^{\prime }}^{\sigma }(r)$ and $%
c_{nlm,n^{\prime }l^{\prime }m^{\prime }}^{\sigma }(r)$ in Eq. (\ref{e213})
are given in Ref. \cite{zhou2005}.

To calculate electron spin-orbital, the Legendre generalized pseudospectral
(LGPS) method \cite{Wang94} is used to discretize the radial KS equation (%
\ref{e205}). This method associated with an appropriate mapping technique
can overcome difficulties due to singularity at $r=0$ and long-tail at large 
$r$ of Coulomb interaction and thus provides a very effective and efficient
numerical algorithm for high-precision solution of KS equation. Using the
electron spin-orbitals of an electronic configuration, a single Slater
determinant for a specific electronic state is constructed and its total
energy calculated. The total energy is a sum of non-interacting
kinetic-energy $E_{k}$, external-field energy $E_{ext}$, Hartree energy $%
E_{H}$, exchange energy $E_{x}$, and correlation energy $E_{c}$. The values
of $E_{k}$, $E_{ext}$, $E_{H}$, and $E_{x}$ are evaluated by%
\begin{equation}
E_{k}=\sum_{\sigma =\alpha }^{\beta }\sum_{nl}^{\nu _{\sigma }}w_{nl\sigma
}\int R_{nl\sigma }\left( r\right) \left( -\frac{1}{2}\frac{d^{2}}{dr^{2}}+%
\frac{l(l+1)}{2r^{2}}\right) R_{nl\sigma }\left( r\right) dr,  \label{e216}
\end{equation}%
\begin{equation}
E_{ext}=4\pi \int v_{ext}\left( r\right) \rho \left( r\right) r^{2}dr,
\label{e217}
\end{equation}%
\begin{equation}
E_{H}=\frac{1}{2}\sum\limits_{\Pi }\eta _{lm,l^{\prime }m^{\prime
}}^{k}F_{nl\sigma ,n^{\prime }l^{\prime }\sigma ^{\prime }}^{k},
\label{e218}
\end{equation}%
and%
\begin{equation}
E_{x}=-\frac{1}{2}\sum\limits_{\Pi }\lambda _{lm,l^{\prime }m^{\prime
}}^{k}G_{nl\sigma ,n^{\prime }l^{\prime }\sigma ^{\prime }}^{k}\delta
_{\sigma \sigma ^{\prime }},  \label{e221}
\end{equation}%
where, $\Pi $ represents a collection of all the quantum numbers involved,
the matrix elements $\eta _{lm,l^{\prime }m^{\prime }}^{k}$, $F_{nl\sigma
,n^{\prime }l^{\prime }\sigma ^{\prime }}^{k}$, and $G_{nl\sigma ,n^{\prime
}l^{\prime }\sigma ^{\prime }}^{k}$ are given in Ref. \cite{zhou2005}.

For a multiplet state that can be described completely by a single Slater
determinant, the energy is calculated directly from the single Slater
determinant. For a multiplet state that cannot be represented by a single
determinant, the energy can be calculated by means of Slater's diagonal sum
rule \cite{Slater60}. According to this rule, a sum over
single-Slater-determinant energy $E($D$_{i})$ of determinant D$_{i}$ from an
electron configuration equals to a weighted sum over multiplet energy $E($M$%
_{j})$ of multiplet state M$_{j}$ involved in the same electron
configuration, namely,%
\begin{equation}
\sum_{i}E(\text{D}_{i})=\sum_{j}d_{j}E(\text{M}_{j}),  \label{e223}
\end{equation}%
where, the weight $d_{j}$ is the times that the multiplet state M$_{j}$
appears in all the single Slater determinants. Similar procedures have been
employed in recent excited-state calculations \cite{a39,a14,Pollak97}.

\section{Results and discussion}

The procedure described in the preceding section is extended to calculate
the total energies $\left( E\right) $ and excitation energies $\left( \Delta
E\right) $ of inner-shell excited states of closed-shell atomic systems: Be,
B$^{+}$, Ne, and Mg. In the calculations, the correlation effect, which is
characterized by the correlation potential $v_{c\sigma }\left( r\right) $
and correlation energy $E_{c}$, is taken into account through the
correlation potentials and energy functionals of Perdew and Wang (PW) \cite%
{a21} and of Lee, Yang, and Parr (LYP) \cite{a38}, respectively. The results
obtained with these two correlation energy functionals are listed in columns
PW and LYP in the following tables, respectively. For simplicity,
henceforward we use an abbreviation $\left( nl\right) ^{-1}\left( n^{\prime
}l^{\prime }\right) $ to represent an electronic configuration for an
electron in inner-shell $\left( nl\right) $ being excited to sub-shell $%
\left( n^{\prime }l^{\prime }\right) $, unless otherwise specified. For
instance, an abbreviation $1s^{-1}2p$ represents an electronic configuration
of inner-shell excitation $1s2s^{2}2p$ of Be.

\subsection{Be}

In TABLE \ref{T-1} we present the total energies and excitation energies
from our calculations for inner-shell excited states $1s^{-1}np$ $^{1,3}P$ ($%
n=2\thicksim 8$) of Be. For comparison we also list in this table the
theoretical results of density work-functional approach (WF) \cite{Roy97b},
saddle-point complex-rotation approximation (SPCR) \cite{Lin01,Lin02}, and
R-matrix method in the close-coupling approximation (RMCC) \cite{Voky92} and
experimental results (Exp.) \cite%
{Jimenez-Mier99,Caldwell90,Jannitti87,Rodbro79}. For the total energies, the
maximum relative discrepancies of our PW and LYP results are 0.10\% and
0.25\% to the WF results, and 0.27\% and 0.39\% to the SPCR results. For
excitation energies, the maximum relative deviations of our PW and LYP
results to the experimental results are 0.37\% and 0.90\%, respectively,
while the maximum relative discrepancies of the WF, SPCR, and RMCC results
to the experimental results are 0.41\%, 0.05\%, and 0.14\%, respectively.
This demonstrates that both the total energies and excitation energies from
our calculations agree well with the experimental and other theoretical
results. On the other hands, due to overestimation of LYP energy functional
to correlation energies of atomic systems with smaller $Z$ \cite%
{a38,zhou2005}, the LYP results are a little bit worse than the PW results.
It is also shown that our PW results for excitation energy are a little bit
better than the WF results.

\begingroup\squeezetable

%TCIMACRO{\TeXButton{B}{\begin{table}[htbp] \centering}}%
%BeginExpansion
\begin{table*}[htbp] \centering%
%EndExpansion
%TCIMACRO{%
%\TeXButton{T-1}{\caption{Total energies ($E$) and excitation energies ($\Delta E$) of inner-shell 
%excited states $1s^{-1}np$ $^{1,3}P$ ($n=2\thicksim 8$) of Be. 
%The ground state energies obtained from calculation with PW and LYP correlation 
%potentials and energy functionals are $-14.6575$ (a.u.) and $-14.6686$ (a.u.), 
%respectively. Here 1 a.u.=27.2116 eV is used.\label{T-1}}}}%
%BeginExpansion
\caption{Total energies ($E$) and excitation energies ($\Delta E$) of inner-shell 
excited states $1s^{-1}np$ $^{1,3}P$ ($n=2\thicksim 8$) of Be. 
The ground state energies obtained from calculation with PW and LYP correlation 
potentials and energy functionals are $-14.6575$ (a.u.) and $-14.6686$ (a.u.), 
respectively. Here 1 a.u.=27.2116 eV is used.\label{T-1}}%
%EndExpansion
\begin{tabular}{ccccccccccccccccccccccccc}
\hline\hline
& \ \ \ \  &  &  & $-E$ & \ \  & (a.u.) &  &  & \ \ \ \  &  &  &  & \ \  & $%
\Delta E$ &  &  &  & (eV) & \ \  &  &  &  &  &  \\ \cline{3-9}\cline{11-25}
States &  & Present &  & work &  & other &  & theory &  & Present &  & work
&  & Other &  &  &  & theory &  & \multicolumn{1}{r}{} &  & 
\multicolumn{1}{l}{Exp.} &  & \multicolumn{1}{l}{} \\ 
\cline{3-5}\cline{7-9}\cline{11-13}\cline{15-19}\cline{21-25}
&  & PW$^{a}$ &  & LYP$^{b}$ &  & WF$^{c}$ &  & SPCR$^{d}$ &  & PW$^{a}$ & 
& LYP$^{b}$ &  & WF$^{c}$ &  & SPCR$^{d}$ &  & RMCC$^{e}$ &  & JSCK$^{f}$ & 
& JNT$^{g}$ &  & RBB$^{h}$ \\ \hline
\multicolumn{1}{r}{$1s^{-1}2p$ $^{3}P$} &  & $10.4526$ &  & $10.4362$ &  & 
10.4628 &  & 10.4654 &  & 114.4221 &  & 115.1704 &  & 114.4304 &  &  &  &  & 
&  &  &  &  & 114.2 \\ 
\multicolumn{1}{r}{$\ \ \ \ \ \ \ \ \ \ \ ^{1}P$} &  & $10.4117$ &  & $%
10.4163$ &  & 10.4146 &  & 10.4209 &  & 115.5350 &  & 115.7119 &  & 115.7420
&  & 115.513 &  & 115.66 &  & 115.49 &  &  &  &  \\ 
\multicolumn{1}{r}{$1s^{-1}3p$ $^{3}P$} &  & $10.1843$ &  & $10.1703$ &  & 
10.1942 &  &  &  & 121.7229 &  & 122.4059 &  & 121.7395 &  &  &  &  &  &  & 
&  &  &  \\ 
\multicolumn{1}{r}{$\ \ \ \ \ \ \ \ \ \ \ ^{1}P$} &  & $10.1797$ &  & $%
10.1671$ &  & 10.1882 &  & 10.2073 &  & 121.8481 &  & 122.4930 &  & 121.9028
&  & 121.420 &  & 121.49 &  & 121.42 &  & 121.4 &  &  \\ 
\multicolumn{1}{r}{$1s^{-1}4p$ $^{3}P$} &  & $10.1410$ &  & $10.1290$ &  & 
10.1504 &  &  &  & 122.9012 &  & 123.5309 &  & 122.9314 &  &  &  &  &  &  & 
&  &  &  \\ 
\multicolumn{1}{r}{$\ \ \ \ \ \ \ \ \ \ \ ^{1}P$} &  & $10.1392$ &  & $%
10.1276$ &  & 10.1480 &  & 10.1662 &  & 122.9502 &  & 123.5690 &  & 122.9967
&  & 122.537 &  & 122.63 &  & 122.52 &  & 122.5 &  &  \\ 
\multicolumn{1}{r}{$1s^{-1}5p$ $^{3}P$} &  & $10.1239$ &  & $10.1123$ &  & 
10.1331 &  &  &  & 123.3665 &  & 123.9848 &  & 123.4021 &  &  &  &  &  &  & 
&  &  &  \\ 
\multicolumn{1}{r}{$\ \ \ \ \ \ \ \ \ \ \ ^{1}P$} &  & $10.1229$ &  & $%
10.1116$ &  & 10.1319 &  & 10.1495 &  & 123.3937 &  & 124.0044 &  & 123.4348
&  & 122.992 &  & 123.08 &  & 122.96 &  & 123.0 &  &  \\ 
\multicolumn{1}{r}{$1s^{-1}6p$ $^{3}P$} &  & $10.1152$ &  & $10.1039$ &  & 
&  &  &  & $123.6033$ &  & $124.2125$ &  &  &  &  &  &  &  &  &  &  &  &  \\ 
\multicolumn{1}{r}{$\ \ \ \ \ \ \ \ \ \ \ ^{1}P$} &  & $10.1146$ &  & $%
10.1035$ &  &  &  & 10.1412 &  & $123.6196$ &  & $124.2240$ &  &  &  & 
123.219 &  &  &  & 123.16 &  &  &  &  \\ 
\multicolumn{1}{r}{$1s^{-1}7p$ $^{3}P$} &  & $10.1102$ &  & $10.0991$ &  & 
&  &  &  & $123.7393$ &  & $124.3431$ &  &  &  &  &  &  &  &  &  &  &  &  \\ 
\multicolumn{1}{r}{$\ \ \ \ \ \ \ \ \ \ \ ^{1}P$} &  & $10.1098$ &  & $%
10.0988$ &  &  &  &  &  & $123.7502$ &  & $124.3508$ &  &  &  &  &  &  &  & 
&  &  &  &  \\ 
\multicolumn{1}{r}{$1s^{-1}8p$ $^{3}P$} &  & $10.1070$ &  & $10.0961$ &  & 
&  &  &  & $123.8264$ &  & $124.4250$ &  &  &  &  &  &  &  &  &  &  &  &  \\ 
\multicolumn{1}{r}{$\ \ \ \ \ \ \ \ \ \ \ ^{1}P$} &  & $10.1068$ &  & $%
10.0959$ &  &  &  &  &  & $123.8318$ &  & $124.4299$ &  &  &  &  &  &  &  & 
&  &  &  &  \\ \hline\hline
\end{tabular}

$^{a}$PW results, $^{b}$LYP results, $^{c}$\cite{Roy97b}, $^{d}$\cite%
{Lin01,Lin02}, $^{e}$\cite{Voky92}, $^{f}$\cite{Jimenez-Mier99,Caldwell90}, $%
^{g}$\cite{Jannitti87}, and $^{h}$\cite{Rodbro79}.%
%TCIMACRO{\TeXButton{E}{\end{table}}}%
%BeginExpansion
\end{table*}%
%EndExpansion
\endgroup

\subsection{B$^{+}$}

To explore the feasibility of the approach to inner-shell excitation of
atomic ions, we also apply the procedure to inner-shell excited-state
calculation of B$^{+}$. The total energies and excitation energies of
inner-shell excited states $1s^{-1}np$ $^{1,3}P$\ ($n=2\thicksim 8$) are
given in TABLE \ref{T-2} along with theoretical results of Dirac-Fock method
(DF)\ \cite{Lynam92} and available experimental results \cite{Lynam92}. For
excitation energies, the relative deviations of our PW and LYP results are
less than 0.32\% and 0.64\% to the DF results, and less than 0.29\% and
0.60\% to the experimental results. This demonstrates again that our results
are in good agreement with both the experimental and other theoretical
results, and the PW results are a little bit more accurate than the LYP
results for this atomic ion having smaller $Z$.

\begingroup\squeezetable

%TCIMACRO{\TeXButton{B}{\begin{table}[htbp] \centering}}%
%BeginExpansion
\begin{table*}[htbp] \centering%
%EndExpansion
%TCIMACRO{%
%\TeXButton{T-2}{\caption{Total energies ($E$) and excitation energies ($\Delta E$) of inner-shell 
%excited states $1s^{-1}np$ $^{1,3}P$ ($n=2\thicksim 8$) of B$^{+}$. 
%The ground state energies obtained from calculation with PW and LYP correlation 
%potentials and energy functionals are $-24.3284$ (a.u.) and $-24.3432$ (a.u.), 
%respectively.\label{T-2}}}}%
%BeginExpansion
\caption{Total energies ($E$) and excitation energies ($\Delta E$) of inner-shell 
excited states $1s^{-1}np$ $^{1,3}P$ ($n=2\thicksim 8$) of B$^{+}$. 
The ground state energies obtained from calculation with PW and LYP correlation 
potentials and energy functionals are $-24.3284$ (a.u.) and $-24.3432$ (a.u.), 
respectively.\label{T-2}}%
%EndExpansion
\begin{tabular}{ccccccccccccc}
\hline\hline
& \ \ \ \ \ \ \  & $-E$ & \  & (a.u.) & \ \ \ \ \ \ \  &  & \  & $\Delta E$
& \ \ \ \  & (eV) & \ \ \ \  &  \\ \cline{3-5}\cline{7-13}
States &  & Present &  & work &  & Present &  & work &  & Other theory &  & 
Exp. \\ \cline{3-5}\cline{7-9}\cline{11-11}\cline{13-13}
&  & PW$^{a}$ &  & LYP$^{b}$ &  & PW$^{a}$ &  & LYP$^{b}$ &  & DF$^{c}$ &  & 
LCC$^{d}$ \\ \hline
\multicolumn{1}{r}{$1s^{-1}2p$ $^{3}P$} &  & \multicolumn{1}{l}{$17.2559$} & 
& \multicolumn{1}{l}{$17.2412$} &  & \multicolumn{1}{l}{$192.4546$} &  & 
\multicolumn{1}{l}{$193.2571$} & \multicolumn{1}{l}{} & 192.460 & 
\multicolumn{1}{l}{} &  \\ 
\multicolumn{1}{r}{$\ \ \ \ \ \ \ \ \ \ \ ^{1}P$} &  & \multicolumn{1}{l}{$%
17.1837$} &  & \multicolumn{1}{l}{$17.1977$} &  & \multicolumn{1}{l}{$%
194.4187$} &  & \multicolumn{1}{l}{$194.4417$} & \multicolumn{1}{l}{} & 
194.394 & \multicolumn{1}{l}{} & 194.39 \\ 
\multicolumn{1}{r}{$1s^{-1}3p$ $^{3}P$} &  & \multicolumn{1}{l}{$16.5968$} & 
& \multicolumn{1}{l}{$16.5858$} &  & \multicolumn{1}{l}{$210.3878$} &  & 
\multicolumn{1}{l}{$211.0921$} & \multicolumn{1}{l}{} & 209.850 & 
\multicolumn{1}{l}{} &  \\ 
\multicolumn{1}{r}{$\ \ \ \ \ \ \ \ \ \ \ ^{1}P$} &  & \multicolumn{1}{l}{$%
16.5861$} &  & \multicolumn{1}{l}{$16.5799$} &  & \multicolumn{1}{l}{$%
210.6797$} &  & \multicolumn{1}{l}{$211.2534$} & \multicolumn{1}{l}{} & 
210.125 & \multicolumn{1}{l}{} & 210.14 \\ 
\multicolumn{1}{r}{$1s^{-1}4p$ $^{3}P$} &  & \multicolumn{1}{l}{$16.4550$} & 
& \multicolumn{1}{l}{$16.4444$} &  & \multicolumn{1}{l}{$214.2478$} &  & 
\multicolumn{1}{l}{$214.9409$} & \multicolumn{1}{l}{} & 213.611 & 
\multicolumn{1}{l}{} &  \\ 
\multicolumn{1}{r}{$\ \ \ \ \ \ \ \ \ \ \ ^{1}P$} &  & \multicolumn{1}{l}{$%
16.4509$} &  & \multicolumn{1}{l}{$16.4422$} &  & \multicolumn{1}{l}{$%
214.3599$} &  & \multicolumn{1}{l}{$214.9983$} & \multicolumn{1}{l}{} & 
213.715 & \multicolumn{1}{l}{} & 213.76 \\ 
\multicolumn{1}{r}{$1s^{-1}5p$ $^{3}P$} &  & \multicolumn{1}{l}{$16.3959$} & 
& \multicolumn{1}{l}{$16.3851$} &  & \multicolumn{1}{l}{$215.8549$} &  & 
\multicolumn{1}{l}{$216.5545$} & \multicolumn{1}{l}{} & 215.189 & 
\multicolumn{1}{l}{} &  \\ 
\multicolumn{1}{r}{$\ \ \ \ \ \ \ \ \ \ \ ^{1}P$} &  & \multicolumn{1}{l}{$%
16.3939$} &  & \multicolumn{1}{l}{$16.3841$} &  & \multicolumn{1}{l}{$%
215.9099$} &  & \multicolumn{1}{l}{$216.5796$} & \multicolumn{1}{l}{} & 
215.237 & \multicolumn{1}{l}{} & 215.30 \\ 
\multicolumn{1}{r}{$1s^{-1}6p$ $^{3}P$} &  & \multicolumn{1}{l}{$16.3654$} & 
& \multicolumn{1}{l}{$16.3546$} &  & \multicolumn{1}{l}{$216.6860$} &  & 
\multicolumn{1}{l}{$217.3845$} & \multicolumn{1}{l}{} & 215.999 & 
\multicolumn{1}{l}{} &  \\ 
\multicolumn{1}{r}{$\ \ \ \ \ \ \ \ \ \ \ ^{1}P$} &  & \multicolumn{1}{l}{$%
16.3642$} &  & \multicolumn{1}{l}{$16.3541$} &  & \multicolumn{1}{l}{$%
216.7170$} &  & \multicolumn{1}{l}{$217.3970$} & \multicolumn{1}{l}{} & 
216.028 & \multicolumn{1}{l}{} & 216.10 \\ 
\multicolumn{1}{r}{$1s^{-1}7p$ $^{3}P$} &  & \multicolumn{1}{l}{$16.3477$} & 
& \multicolumn{1}{l}{$16.3368$} &  & \multicolumn{1}{l}{$217.1684$} &  & 
\multicolumn{1}{l}{$217.8672$} & \multicolumn{1}{l}{} &  & 
\multicolumn{1}{l}{} &  \\ 
\multicolumn{1}{r}{$\ \ \ \ \ \ \ \ \ \ \ ^{1}P$} &  & \multicolumn{1}{l}{$%
16.3469$} &  & \multicolumn{1}{l}{$16.3365$} &  & \multicolumn{1}{l}{$%
217.1886$} &  & \multicolumn{1}{l}{$217.8745$} & \multicolumn{1}{l}{} &  & 
\multicolumn{1}{l}{} &  \\ 
\multicolumn{1}{r}{$1s^{-1}8p$ $^{3}P$} &  & \multicolumn{1}{l}{$16.3362$} & 
& \multicolumn{1}{l}{$16.3256$} &  & \multicolumn{1}{l}{$217.4803$} &  & 
\multicolumn{1}{l}{$218.1725$} & \multicolumn{1}{l}{} &  & 
\multicolumn{1}{l}{} &  \\ 
\multicolumn{1}{r}{$\ \ \ \ \ \ \ \ \ \ \ ^{1}P$} &  & \multicolumn{1}{l}{$%
16.3359$} &  & \multicolumn{1}{l}{$16.3254$} &  & \multicolumn{1}{l}{$%
217.4895$} &  & \multicolumn{1}{l}{$218.1767$} & \multicolumn{1}{l}{} &  & 
\multicolumn{1}{l}{} &  \\ \hline\hline
\end{tabular}

$^{a}$PW results, $^{b}$LYP results, $^{c}$\cite{Lynam92}, and $^{d}$\cite%
{Lynam92}$.$%
%TCIMACRO{\TeXButton{E}{\end{table}}}%
%BeginExpansion
\end{table*}%
%EndExpansion
\endgroup

\subsection{Ne}

We present in TABLE \ref{T-3} and TABLE \ref{T-4} the total energies and
excitation energies from our calculations for inner-shell excited states $%
1s^{-1}ns$ $^{1,3}S$ $\left( n=3\sim 8\right) $ and $1s^{-1}np$ $^{1,3}P$ $%
\left( n=3\sim 8\right) $ of Ne, respectively. We also present in TABLE \ref%
{T-4} the total energy of inner-shell excited state $1s2s^{2}2p^{6}$ $^{2}S$
of Ne$^{+}$ and ionization energy of an inner-shell $1s$ electron of Ne. For
comparison we also show in these tables the theoretical results of density
work-functional approach (WF) \cite{Roy97b}, configuration-interaction model
(CI) \cite{Schroter99}, and Hartree-Fock method (HF) \cite{Sewell65}, and
experimental results (Exp.) \cite%
{Hitchcock80,Sodhi84,Domke92,Avaldi95,Liefielf65,Wuilleumier70,Codling67,Simpson64}%
.

For the total energies of excited states $1s^{-1}ns$ $^{1,3}S$ $\left(
n=3\sim 8\right) $ given in TABLE \ref{T-3}, the relative deviations of our
PW and LYP results to the WF results are not more than 0.03\% and 0.02\%,
respectively. This demonstrates that the PW energy functional has almost the
same precision as the LYP energy functional in calculation of total energies
of these inner-shell excited states. For excitation energies, the relative
discrepancies of our PW and LYP results to the experimental result are less
than 0.13\% and 0.02\%, respectively. This indicates that the LYP results
for the excitation energy are better than the PW results. Since the total
energies from calculation with PW energy functional are very close to those
with LYP energy functional, the larger discrepancy of the PW results for
excitation energy mainly comes from the ground-state energy, which is
-128.8952 (a.u.) from the calculation with PW energy functional. This value
is different from that with LYP energy functional -128.9331 (a.u.) \cite%
{zhou2005} and that obtained from Hartree-Fock energy \cite{Veillard68} plus
correlation energy \cite{a25} -128.937 (a.u.). In addition, the maximum
relative discrepancies of the excitation energies from WF calculation and CI
calculation are 0.08\% and 0.03\% to the experimental results, respectively.
This illustrates that our LYP results are very close to the CI results and
better than the WF results.

For the total energies of inner-shell excited states $1s^{-1}np$ $^{1,3}P$ $%
\left( n=3\sim 8\right) $ given in TABLE \ref{T-4}, the relative deviations
of our PW and LYP results to the WF results are not more than 0.02\% and
0.02\%, respectively. This implies that the PW energy functional has the
same precision as the LYP energy functional in the total energy calculation
of these states. For the excitation energies, the maximum relative
discrepancies of our PW and LYP results to the experimental results are
0.14\% and 0.02\%, while the maximum relative deviations of the WF and CI
results to the experimental results are 0.07\% and 0.02\%, respectively.
This demonstrates that the LYP results for excitation energy are a little
bit more accurate than the PW results for these inner-shell excited states.
It is also shown that the LYP results are again very close to the CI results
and a little bit better than the WF results.

\begingroup\squeezetable

%TCIMACRO{\TeXButton{B}{\begin{table}[htbp] \centering}}%
%BeginExpansion
\begin{table*}[htbp] \centering%
%EndExpansion
%TCIMACRO{%
%\TeXButton{T-3}{\caption{Total energies ($E$) and excitation energies ($\Delta E$) of inner-shell
%excited states $1s^{-1}ns$ $^{1,3}S$ ($n=3\thicksim 8$) of Ne.
%The ground state energies from calculations with PW and LYP correlation 
%potentials and energy functionals 
%are -128.8952 (a.u.) and -128.9331 (a.u.), respectively.\label{T-3}}}}%
%BeginExpansion
\caption{Total energies ($E$) and excitation energies ($\Delta E$) of inner-shell
excited states $1s^{-1}ns$ $^{1,3}S$ ($n=3\thicksim 8$) of Ne.
The ground state energies from calculations with PW and LYP correlation 
potentials and energy functionals 
are -128.8952 (a.u.) and -128.9331 (a.u.), respectively.\label{T-3}}%
%EndExpansion
\begin{tabular}{ccccccccccccccccccc}
\hline\hline
& \ \ \ \ \  & $-E$ &  &  &  & (a.u.) & \ \ \ \ \  &  &  & $\Delta E$ & \ \ 
&  &  & (eV) & \ \  &  &  &  \\ \cline{3-7}\cline{9-19}
States &  & Present &  & work &  & Other theory &  & Present &  & work &  & 
Other &  & theory &  &  & Exp. &  \\ 
\cline{3-5}\cline{7-7}\cline{9-11}\cline{13-15}\cline{17-19}
&  & PW$^{a}$ &  & LYP$^{b}$ &  & WF$^{c}$ &  & PW$^{a}$ &  & LYP$^{b}$ &  & 
WF$^{c}$ &  & CI$^{d}$ &  & HB$^{e}$ &  & SAC$^{f}$ \\ \hline
\multicolumn{1}{r}{$1s^{-1}3s\text{ }^{3}S$} &  & $97.1443$ &  & $97.1495$ & 
& $97.1729$ &  & $863.9920$ &  & $864.8826$ &  & $864.3917$ &  &  &  &  &  & 
\\ 
\multicolumn{1}{r}{$^{1}S$} &  & $97.1381$ &  & $97.1411$ &  & $97.1631$ & 
& $864.1601$ &  & $865.1112$ &  & $864.6583$ &  & $865.37$ &  & 865.1 &  & 
865.32 \\ 
\multicolumn{1}{r}{$1s^{-1}4s\text{ }^{3}S$} &  & $97.0335$ &  & $97.0348$ & 
&  &  & $867.0081$ &  & $868.0038$ &  &  &  &  &  &  &  &  \\ 
\multicolumn{1}{r}{$^{1}S$} &  & $97.0319$ &  & $97.0326$ &  &  &  & $%
867.0516$ &  & $868.0636$ &  &  &  & $868.21$ &  &  &  &  \\ 
\multicolumn{1}{r}{$1s^{-1}5s\text{ }^{3}S$} &  & $96.9999$ &  & $97.0008$ & 
&  &  & $867.9208$ &  & $868.9290$ &  &  &  &  &  &  &  &  \\ 
\multicolumn{1}{r}{$^{1}S$} &  & $96.9993$ &  & $97.0000$ &  &  &  & $%
867.9382$ &  & $868.9507$ &  &  &  & $869.06$ &  &  &  &  \\ 
\multicolumn{1}{r}{$1s^{-1}6s\text{ }^{3}S$} &  & $96.9853$ &  & $96.9863$ & 
&  &  & $868.3189$ &  & $869.3235$ &  &  &  &  &  &  &  &  \\ 
\multicolumn{1}{r}{$^{1}S$} &  & $96.9850$ &  & $96.9859$ &  &  &  & $%
868.3276$ &  & $869.3344$ &  &  &  & $869.44$ &  &  &  &  \\ 
\multicolumn{1}{r}{$1s^{-1}7s\text{ }^{3}S$} &  & $96.9776$ &  & $96.9788$ & 
&  &  & $868.5279$ &  & $869.5276$ &  &  &  &  &  &  &  &  \\ 
\multicolumn{1}{r}{$^{1}S$} &  & $96.9774$ &  & $96.9786$ &  &  &  & $%
868.5328$ &  & $869.5331$ &  &  &  &  &  &  &  &  \\ 
\multicolumn{1}{r}{$1s^{-1}8s\text{ }^{3}S$} &  & $96.9731$ &  & $96.9744$ & 
&  &  & $868.6522$ &  & $869.6474$ &  &  &  &  &  &  &  &  \\ 
\multicolumn{1}{r}{$^{1}S$} &  & $96.9729$ &  & $96.9743$ &  &  &  & $%
868.6555$ &  & $869.6501$ &  &  &  &  &  &  &  &  \\ \hline\hline
\end{tabular}

$^{a}$PW results, $^{b}$LYP results, $^{c}$\cite{Roy97b}, $^{d}$\cite%
{Schroter99}, $^{e}$\cite{Hitchcock80,Sodhi84}, and $^{f}$\cite{Schroter99}.%
%TCIMACRO{\TeXButton{E}{\end{table}}}%
%BeginExpansion
\end{table*}%
%EndExpansion
\endgroup

\bigskip

\begingroup\squeezetable

%TCIMACRO{\TeXButton{B}{\begin{table}[htbp] \centering}}%
%BeginExpansion
\begin{table*}[htbp] \centering%
%EndExpansion
%TCIMACRO{%
%\TeXButton{T-4}{\caption{Total energies ($E$) and excitation energies ($\Delta E$) of inner-shell
%excited states $1s^{-1}np$ $^{1,3}P$ ($n=3\thicksim 8$) of Ne and 
%$1s2s^22p^6$ of Ne$^+$.\label{T-4}}}}%
%BeginExpansion
\caption{Total energies ($E$) and excitation energies ($\Delta E$) of inner-shell
excited states $1s^{-1}np$ $^{1,3}P$ ($n=3\thicksim 8$) of Ne and 
$1s2s^22p^6$ of Ne$^+$.\label{T-4}}%
%EndExpansion
\begin{tabular}{ccccccccccccccccccccc}
\hline\hline
& \ \ \ \ \  & $-E$ &  &  & \ \  & (a.u.) & \ \ \ \ \  &  &  & $\Delta E$ & 
\ \  &  &  &  & \ \  &  &  & (eV) &  &  \\ \cline{3-7}\cline{9-21}
States &  & Present &  & work &  & Other theory &  & Present &  & work &  & 
Other &  & theory &  &  &  & Exp. &  &  \\ 
\cline{3-5}\cline{7-7}\cline{9-11}\cline{13-15}\cline{17-21}
&  & PW$^{a}$ &  & LYP$^{b}$ &  & WF$^{c}$ &  & PW$^{a}$ &  & LYP$^{b}$ &  & 
WF$^{c}$ &  & CI$^{d}$ &  & HB$^{e}$ &  & W$^{f}$ &  & ADC$^{g}$ \\ \hline
\multicolumn{1}{r}{$1s^{-1}3p\text{ }^{3}P$} &  & $97.0762$ &  & $97.0766$ & 
& $97.0982$ &  & $865.8456$ &  & $866.8663$ &  & $866.4244$ &  &  &  &  &  & 
&  &  \\ 
\multicolumn{1}{r}{$^{1}P$} &  & $97.0743$ &  & $97.0736$ &  & $97.0950$ & 
& $865.8984$ &  & $866.9480$ &  & $866.5115$ &  & $867.18$ &  & $867.05$ & 
& $867.13$ &  & 867.12 \\ 
\multicolumn{1}{r}{$1s^{-1}4p\text{ }^{3}P$} &  & $97.0147$ &  & $97.0151$ & 
& $97.0330$ &  & $867.5194$ &  & $868.5398$ &  & $868.1986$ &  &  &  &  &  & 
&  &  \\ 
\multicolumn{1}{r}{$^{1}P$} &  & $97.0140$ &  & $97.0141$ &  & $97.0318$ & 
& $867.5377$ &  & $868.5671$ &  & $868.2312$ &  & $868.70$ &  & $868.68$ & 
& $868.77$ &  & 868.69 \\ 
\multicolumn{1}{r}{$1s^{-1}5p\text{ }^{3}P$} &  & $96.9914$ &  & $96.9919$ & 
& $97.0098$ &  & $868.1540$ &  & $869.1712$ &  & $868.8299$ &  &  &  &  &  & 
&  &  \\ 
\multicolumn{1}{r}{$^{1}P$} &  & $96.9910$ &  & $96.9915$ &  & $97.0095$ & 
& $868.1627$ &  & $869.1820$ &  & $868.8462$ &  & $869.32$ &  & $869.23$ & 
& $869.37$ &  & 869.27 \\ 
\multicolumn{1}{r}{$1s^{-1}6p\text{ }^{3}P$} &  & $96.9806$ &  & $96.9814$ & 
& $96.9900$ &  & $868.4457$ &  & $869.4569$ &  & $869.1238$ &  &  &  &  &  & 
&  &  \\ 
\multicolumn{1}{r}{$^{1}P$} &  & $96.9805$ &  & $96.9812$ &  & $96.9988$ & 
& $868.4506$ &  & $869.4623$ &  & $869.1347$ &  & $869.58$ &  & $869.63$ & 
& $869.65$ &  & 869.56 \\ 
\multicolumn{1}{r}{$1s^{-1}7p\text{ }^{3}P$} &  & $96.9747$ &  & $96.9756$ & 
&  &  & $868.6071$ &  & $869.6147$ &  &  &  &  &  &  &  &  &  &  \\ 
\multicolumn{1}{r}{$^{1}P$} &  & $96.9746$ &  & $96.9755$ &  &  &  & $%
868.6092$ &  & $869.6174$ &  &  &  & $869.79$ &  &  &  &  &  & 869.73 \\ 
\multicolumn{1}{r}{$1s^{-1}8p\text{ }^{3}P$} &  & $96.9710$ &  & $96.9721$ & 
&  &  & $868.7088$ &  & $869.7099$ &  &  &  &  &  &  &  &  &  &  \\ 
\multicolumn{1}{r}{$^{1}P$} &  & $96.9710$ &  & $96.9720$ &  &  &  & $%
868.7088$ &  & $869.7127$ &  &  &  & $869.87$ &  &  &  &  &  &  \\ 
\multicolumn{1}{r}{} &  &  &  &  &  &  &  &  &  &  &  &  &  &  &  &  &  &  & 
&  \\ 
\multicolumn{1}{r}{$1s2s^{2}2p^{6}$ $^{2}S$} &  & $96.9612$ &  & $96.9629$ & 
&  &  & $868.9741$ &  & $869.9603$ &  & $869.6898$ &  & $870.15$ &  & $%
870.10 $ &  & $870.17$ &  & 870.17 \\ \hline\hline
\end{tabular}

$^{a}$PW results, $^{b}$LYP results, $^{c}$\cite{Roy97b}, $^{d}$\cite%
{Schroter99} (see also \cite{Coreno99}), $^{e}$\cite{Hitchcock80,Sodhi84}, $%
^{f}$\cite{Wuilleumier70}, and $^{g}$\cite{Coreno99} (see also \cite%
{Schroter99,Avaldi95}).%
%TCIMACRO{\TeXButton{E}{\end{table}}}%
%BeginExpansion
\end{table*}%
%EndExpansion
\endgroup

\subsection{Mg}

For Mg, we have computed the total energies and excitation energies for
inner-shell excited states $2p^{-1}ns$ $^{1,3}P$ $\left( n=4\sim 8\right) $, 
$2s^{-1}np$ $^{1,3}P$ $\left( n=3\sim 8\right) $, $2s^{-1}ns$ $^{1,3}S$ $%
\left( n=4\sim 8\right) $, $1s^{-1}np$ $^{1,3}P$ $\left( n=3\sim 8\right) $,
and $1s^{-1}ns$ $^{1,3}S$ $\left( n=4\sim 8\right) $. The results are shown
in TABLE \ref{T-5} to TABLE \ref{T-9}, respectively. For comparison, the
theoretical results from configuration-interaction calculation with improved
and optimized orbitals (CIIOO) \cite{Martins97} and experimental results
(Exp.) \cite{Martin80,Newsom71} are also shown in these tables.

For inner-shell excitation of Mg, the excited states involving in excitation
of a $2p$ electron, $2p^{-1}ns$ $^{1,3}P$, are the only ones that
experimental excitation energies are available, as shown in TABLE \ref{T-5}.
For these states, the excitation energies from our calculation are in good
agreement with the experimental results. The relative deviations of our PW
and LYP results to the experimental results are not more than 0.38\% and
0.36\%, respectively. Thus the PW energy functional has the same precision
as the LYP energy functional in the excitation energy calculation of these
inner-shell excited states. Apart from an excited state $2p^{-1}4s$ $^{3}P$
our results also agree well with the CIIOO results. The maximum relative
discrepancies of our PW and LYP results to the CIIOO results are 0.58\% and
0.60\%, respectively. The larger discrepancies of our results to the CIIOO
results for the state $2p^{-1}4s$ $^{3}P$ are caused by the fact that the
excitation energy from the CIIOO calculation is too small. It is shown from
TABLE \ref{T-5} that the CIIOO result is much smaller than the experimental
data while our excitation energies are very close to the experimental
results.

For inner-shell excited states relevant to excitation of a $2s$ electron,
the excitation energies from our calculations with both PW and LYP energy
functionals are larger than the CIIOO results, as shown in TABLE \ref{T-6}
and TABLE \ref{T-7}. For the excited states $2s^{-1}np$ $^{3}P$, except an
excited state $2s^{-1}3p$ $^{3}P$, the relative deviations of our PW and LYP
results to the CIIOO results are less than 0.70\% and 0.62\%, respectively.
For the excited state $2s^{-1}3p$ $^{3}P$ our PW and LYP results are much
larger than the CIIOO result. For the excited states $2s^{-1}ns$ $^{3}S$,
apart from an excited state $2s^{-1}4s$ $^{3}S$, the relative discrepancies
of our PW and LYP results to the CIIOO results are not more than 0.41\% and
0.25\%, respectively. For the excited state $2s^{-1}4s$ $^{3}S$ our PW and
LYP results are again much larger than the CIIOO results.

For excitation energies of excited states $1s^{-1}np$ $^{3}P$ and $1s^{-1}ns$
$^{3}S$\ given in TABLE \ref{T-8} and TABLE \ref{T-9}, our PW and LYP
results are smaller and larger than the CIIOO results, respectively. For
excited states $1s^{-1}np$ $^{3}P$ the relative deviations of our PW and LYP
results to the CIIOO results are less than 0.02\% and 0.08\%, respectively.
For excited states $1s^{-1}ns$ $^{3}S$, apart from a state $1s^{-1}4s$ $^{3}S
$, the relative discrepancies of our PW and LYP results to the CIIOO results
are not more than 0.04\% and 0.03\%, respectively. For the state $1s^{-1}4s$ 
$^{3}S$ our LYP result is larger than the CIIOO result by 0.11\%.

\begingroup\squeezetable

%TCIMACRO{\TeXButton{B}{\begin{table}[htbp] \centering}}%
%BeginExpansion
\begin{table*}[htbp] \centering%
%EndExpansion
%TCIMACRO{%
%\TeXButton{T-5}{\caption{Total energies ($E$) and excitation energies ($\Delta E$) of inner-shell
%excited states $2p^{-1}ns$ $^{1,3}P$ ($n=4\thicksim 8$) of Mg.
%The ground state energies from calculations with PW and LYP correlation 
%potentials and energy functionals 
%are -200.0204 (a.u.) and -200.0744 (a.u.), respectively.\label{T-5}}}}%
%BeginExpansion
\caption{Total energies ($E$) and excitation energies ($\Delta E$) of inner-shell
excited states $2p^{-1}ns$ $^{1,3}P$ ($n=4\thicksim 8$) of Mg.
The ground state energies from calculations with PW and LYP correlation 
potentials and energy functionals 
are -200.0204 (a.u.) and -200.0744 (a.u.), respectively.\label{T-5}}%
%EndExpansion
\begin{tabular}{ccccccccccccccc}
\hline\hline
& \ \ \ \ \  & $-E$ &  & (a.u.) & \ \ \ \ \  &  &  & $\Delta E$ & \ \  &  & 
\ \  & (eV) &  &  \\ \cline{3-5}\cline{7-15}
States &  & Present &  & work &  & Present &  & work &  & Other theory &  & 
\multicolumn{1}{r}{} & Exp. & \multicolumn{1}{l}{} \\ 
\cline{3-5}\cline{7-9}\cline{11-11}\cline{13-15}
&  & PW$^{a}$ &  & LYP$^{b}$ &  & PW$^{a}$ &  & LYP$^{b}$ &  & CIIOO$^{c}$ & 
& MZ$^{d}$ &  & NE$^{e}$ \\ \hline
\multicolumn{1}{r}{$2p^{-1}4s$ $^{3}P$} &  & $198.0039$ &  & $198.0557$ &  & 
$54.8714$ &  & $54.9323$ &  & 53.72 &  & 54.801 &  & 54.801 \\ 
\multicolumn{1}{r}{$\ \ \ \ \ \ \ \ \ \ \ ^{1}P$} &  & $198.0022$ &  & $%
198.0538$ &  & $54.9190$ &  & $54.9849$ &  &  &  & 55.065 &  & 55.065 \\ 
\multicolumn{1}{r}{$2p^{-1}5s$ $^{3}P$} &  & $197.9509$ &  & $198.0044$ &  & 
$56.3133$ &  & $56.3272$ &  & 55.99 &  & 56.278 &  & 56.280 \\ 
\multicolumn{1}{r}{$\ \ \ \ \ \ \ \ \ \ \ ^{1}P$} &  & $197.9504$ &  & $%
198.0038$ &  & $56.3294$ &  & $56.3457$ &  &  &  & 56.544 &  & 56.545 \\ 
\multicolumn{1}{r}{$2p^{-1}6s$ $^{3}P$} &  & $197.9313$ &  & $197.9852$ &  & 
$56.8469$ &  & $56.8499$ &  & 56.56 &  & 56.777 &  & 56.785 \\ 
\multicolumn{1}{r}{$\ \ \ \ \ \ \ \ \ \ \ ^{1}P$} &  & $197.9312$ &  & $%
197.9849$ &  & $56.8505$ &  & $56.8586$ &  &  &  & 57.039 &  &  \\ 
\multicolumn{1}{r}{$2p^{-1}7s$ $^{3}P$} &  & $197.9217$ &  & $197.9759$ &  & 
$57.1095$ &  & $57.1038$ &  &  &  &  &  &  \\ 
\multicolumn{1}{r}{$\ \ \ \ \ \ \ \ \ \ \ ^{1}P$} &  & $197.9215$ &  & $%
197.9757$ &  & $57.1142$ &  & $57.1087$ &  &  &  & 57.302 &  & 57.305 \\ 
\multicolumn{1}{r}{$2p^{-1}8s$ $^{3}P$} &  & $197.9163$ &  & $197.9707$ &  & 
$57.2573$ &  & $57.2464$ &  &  &  &  &  &  \\ 
\multicolumn{1}{r}{$\ \ \ \ \ \ \ \ \ \ \ ^{1}P$} &  & $197.9161$ &  & $%
197.9706$ &  & $57.2603$ &  & $57.2478$ &  &  &  &  &  & 57.456 \\ 
\hline\hline
\end{tabular}

$^{a}$PW results, $^{b}$LYP results,\ $^{c}$\cite{Martins97}, $^{d}$\cite%
{Martin80}, and $^{e}$\cite{Newsom71}.%
%TCIMACRO{\TeXButton{E}{\end{table}}}%
%BeginExpansion
\end{table*}%
%EndExpansion
\endgroup

\bigskip

\begingroup\squeezetable

%TCIMACRO{\TeXButton{B}{\begin{table}[htbp] \centering}}%
%BeginExpansion
\begin{table}[htbp] \centering%
%EndExpansion
%TCIMACRO{%
%\TeXButton{T-6}{\caption{Total energies ($E$) and excitation energies ($\Delta E$) of inner-shell
%excited states $2s^{-1}np$ $^{1,3}P$ ($n=3\thicksim 8$) of Mg.
%\label{T-6}}}}%
%BeginExpansion
\caption{Total energies ($E$) and excitation energies ($\Delta E$) of inner-shell
excited states $2s^{-1}np$ $^{1,3}P$ ($n=3\thicksim 8$) of Mg.
\label{T-6}}%
%EndExpansion
\begin{tabular}{ccccccccccc}
\hline\hline
&  & $-E$ &  & (a.u.) &  & $\Delta E$ &  &  &  & (eV) \\ 
\cline{3-5}\cline{7-11}
States &  & Present &  & work &  & Present &  & work &  & Other theory \\ 
\cline{3-5}\cline{7-11}
&  & PW$^{a}$ &  & LYP$^{b}$ &  & PW$^{a}$ &  & LYP$^{b}$ &  & CIIOO$^{c}$
\\ \hline
\multicolumn{1}{r}{$2s^{-1}3p$ $^{3}P$} &  & $196.6182$ &  & $196.6721$ &  & 
$92.5785$ &  & $92.5820$ &  & 91.72 \\ 
\multicolumn{1}{r}{$\ \ \ \ \ \ \ \ \ \ \ ^{1}P$} &  & $196.6101$ &  & $%
196.6630$ &  & $92.8005$ &  & $92.8288$ &  &  \\ 
\multicolumn{1}{r}{$2s^{-1}4p$ $^{3}P$} &  & $196.4612$ &  & $196.5181$ &  & 
$96.8526$ &  & $96.7718$ &  & 96.18 \\ 
\multicolumn{1}{r}{$\ \ \ \ \ \ \ \ \ \ \ ^{1}P$} &  & $196.460$ &  & $%
196.5170$ &  & $96.8820$ &  & $96.8026$ &  &  \\ 
\multicolumn{1}{r}{$2s^{-1}5p$ $^{3}P$} &  & $196.4279$ &  & $196.4871$ &  & 
$97.7574$ &  & $97.6164$ &  & 97.11 \\ 
\multicolumn{1}{r}{$\ \ \ \ \ \ \ \ \ \ \ ^{1}P$} &  & $196.4275$ &  & $%
196.4865$ &  & $97.7677$ &  & $97.6317$ &  &  \\ 
\multicolumn{1}{r}{$2s^{-1}6p$ $^{3}P$} &  & $196.4138$ &  & $196.4738$ &  & 
$98.1403$ &  & $97.9775$ &  &  \\ 
\multicolumn{1}{r}{$\ \ \ \ \ \ \ \ \ \ \ ^{1}P$} &  & $196.4136$ &  & $%
196.4735$ &  & $98.1463$ &  & $97.9863$ &  &  \\ 
\multicolumn{1}{r}{$2s^{-1}7p$ $^{3}P$} &  & $196.4065$ &  & $196.4669$ &  & 
$98.3403$ &  & $98.1667$ &  &  \\ 
\multicolumn{1}{r}{$\ \ \ \ \ \ \ \ \ \ \ ^{1}P$} &  & $196.4064$ &  & $%
196.4667$ &  & $98.3441$ &  & $98.1721$ &  &  \\ 
\multicolumn{1}{r}{$2s^{-1}8p$ $^{3}P$} &  & $196.4022$ &  & $196.4628$ &  & 
$98.4581$ &  & $98.2782$ &  &  \\ 
\multicolumn{1}{r}{$\ \ \ \ \ \ \ \ \ \ \ ^{1}P$} &  & $196.4021$ &  & $%
196.4626$ &  & $98.4603$ &  & $98.2820$ &  &  \\ \hline\hline
\end{tabular}

$^{a}$PW results, $^{b}$LYP results, and $^{c}$\cite{Martins97}.%
%TCIMACRO{\TeXButton{E}{\end{table}}}%
%BeginExpansion
\end{table}%
%EndExpansion
\endgroup

\bigskip

\begingroup\squeezetable

%TCIMACRO{\TeXButton{B}{\begin{table}[htbp] \centering}}%
%BeginExpansion
\begin{table}[htbp] \centering%
%EndExpansion
%TCIMACRO{%
%\TeXButton{T-7}{\caption{Total energies ($E$) and excitation energies ($\Delta E$) of inner-shell
%excited states $2s^{-1}ns$ $^{1,3}S$ ($n=4\thicksim 8$) of Mg.
%\label{T-7}}}}%
%BeginExpansion
\caption{Total energies ($E$) and excitation energies ($\Delta E$) of inner-shell
excited states $2s^{-1}ns$ $^{1,3}S$ ($n=4\thicksim 8$) of Mg.
\label{T-7}}%
%EndExpansion
\begin{tabular}{ccccccccccc}
\hline\hline
&  & $-E$ &  & (a.u.) &  & $\Delta E$ &  &  &  & (eV) \\ 
\cline{3-5}\cline{7-11}
States &  & Present &  & work &  & Present &  & work &  & Other theory \\ 
\cline{3-5}\cline{7-9}\cline{11-11}
&  & PW$^{a}$ &  & LYP$^{b}$ &  & PW$^{a}$ &  & LYP$^{b}$ &  & CIIOO$^{c}$
\\ \hline
\multicolumn{1}{r}{$2s^{-1}4s$ $^{3}S$} &  & $196.4919$ &  & $196.5499$ &  & 
$96.0167$ &  & $95.9065$ &  & 94.65 \\ 
\multicolumn{1}{r}{$\ \ \ \ ^{1}S$} &  & $196.4873$ &  & $196.5447$ &  & $%
96.1419$ &  & $96.0502$ &  &  \\ 
\multicolumn{1}{r}{$2s^{-1}5s$ $^{3}S$} &  & $196.4387$ &  & $196.4984$ &  & 
$97.4635$ &  & $97.3079$ &  & 97.07 \\ 
\multicolumn{1}{r}{$\ \ \ \ \ \ \ \ \ \ ^{1}S$} &  & $196.4372$ &  & $%
196.4966$ &  & $97.5060$ &  & $97.3585$ &  &  \\ 
\multicolumn{1}{r}{$2s^{-1}6s$ $^{3}S$} &  & $196.4191$ &  & $196.4792$ &  & 
$97.9985$ &  & $97.8325$ &  & 97.64 \\ 
\multicolumn{1}{r}{$\ \ \ \ \ \ \ \ \ \ ^{1}S$} &  & $196.4183$ &  & $%
196.4783$ &  & $98.0181$ &  & $97.8565$ &  &  \\ 
\multicolumn{1}{r}{$2s^{-1}7s$ $^{3}S$} &  & $196.4094$ &  & $196.4698$ &  & 
$98.2616$ &  & $98.0872$ &  &  \\ 
\multicolumn{1}{r}{$\ \ \ \ \ \ \ \ ^{1}S$} &  & $196.4090$ &  & $196.4693$
&  & $98.2720$ &  & $98.1003$ &  &  \\ 
\multicolumn{1}{r}{$2s^{-1}8s$ $^{3}S$} &  & $196.4039$ &  & $196.4645$ &  & 
$98.4105$ &  & $98.2301$ &  &  \\ 
\multicolumn{1}{r}{$\ \ \ \ \ \ \ ^{1}S$} &  & $196.4037$ &  & $196.4642$ & 
& $98.4176$ &  & $98.2382$ &  &  \\ \hline\hline
\end{tabular}

$^{a}$PW results, $^{b}$LYP results, and $^{c}$\cite{Martins97}.%
%TCIMACRO{\TeXButton{E}{\end{table}}}%
%BeginExpansion
\end{table}%
%EndExpansion
\endgroup

\bigskip

\begingroup\squeezetable

%TCIMACRO{\TeXButton{B}{\begin{table}[htbp] \centering}}%
%BeginExpansion
\begin{table}[htbp] \centering%
%EndExpansion
%TCIMACRO{%
%\TeXButton{T-8}{\caption{Total energies ($E$) and excitation energies ($\Delta E$) of inner-shell
%excited states $1s^{-1}np$ $^{1,3}P$ ($n=3\thicksim 8$) of Mg.
%\label{T-8}}}}%
%BeginExpansion
\caption{Total energies ($E$) and excitation energies ($\Delta E$) of inner-shell
excited states $1s^{-1}np$ $^{1,3}P$ ($n=3\thicksim 8$) of Mg.
\label{T-8}}%
%EndExpansion
\begin{tabular}{ccccccccccc}
\hline\hline
&  & $-E$ &  & (a.u.) &  & $\Delta E$ &  &  &  & (eV) \\ 
\cline{3-5}\cline{7-11}
States &  & Present &  & work &  & Present &  & work &  & Other theory \\ 
\cline{3-5}\cline{7-9}\cline{11-11}
&  & PW$^{a}$ &  & LYP$^{b}$ &  & PW$^{a}$ &  & LYP$^{b}$ &  & CIIOO$^{c}$
\\ \hline
\multicolumn{1}{r}{$1s^{-1}3p$ $^{3}P$} &  & $152.1281$ &  & $152.1433$ &  & 
$1303.2275$ &  & $1304.2822$ &  & 1303.25 \\ 
\multicolumn{1}{r}{$\ \ \ \ \ \ \ \ \ \ \ ^{1}P$} &  & $152.1207$ &  & $%
152.1353$ &  & $1303.4275$ &  & $1304.5002$ &  &  \\ 
\multicolumn{1}{r}{$1s^{-1}4p$ $^{3}P$} &  & $151.9671$ &  & $151.9851$ &  & 
$1307.6080$ &  & $1308.5868$ &  & 1307.86 \\ 
\multicolumn{1}{r}{$\ \ \ \ \ \ \ \ \ \ \ ^{1}P$} &  & $151.9659$ &  & $%
151.9839$ &  & $1307.6412$ &  & $1308.6192$ &  &  \\ 
\multicolumn{1}{r}{$1s^{-1}5p$ $^{3}P$} &  & $151.9334$ &  & $151.9536$ &  & 
$1308.5256$ &  & $1309.4453$ &  & 1308.80 \\ 
\multicolumn{1}{r}{$\ \ \ \ \ \ \ \ \ \ \ ^{1}P$} &  & $151.9329$ &  & $%
151.9533$ &  & $1308.5375$ &  & $1309.4521$ &  &  \\ 
\multicolumn{1}{r}{$1s^{-1}6p$ $^{3}P$} &  & $151.9191$ &  & $151.9401$ &  & 
$1308.9131$ &  & $1309.8108$ &  &  \\ 
\multicolumn{1}{r}{$\ \ \ \ \ \ \ \ \ \ \ ^{1}P$} &  & $151.9189$ &  & $%
151.9399$ &  & $1308.9196$ &  & $1309.8179$ &  &  \\ 
\multicolumn{1}{r}{$1s^{-1}7p$ $^{3}P$} &  & $151.9117$ &  & $151.9331$ &  & 
$1309.1150$ &  & $1310.0018$ &  &  \\ 
\multicolumn{1}{r}{$\ \ \ \ \ \ \ \ \ \ \ ^{1}P$} &  & $151.9116$ &  & $%
151.9329$ &  & $1309.1188$ &  & $1310.0062$ &  &  \\ 
\multicolumn{1}{r}{$1s^{-1}8p$ $^{3}P$} &  & $151.9073$ &  & $151.9290$ &  & 
$1309.2344$ &  & $1310.1145$ &  &  \\ 
\multicolumn{1}{r}{$\ \ \ \ \ \ \ \ \ \ \ ^{1}P$} &  & $151.9072$ &  & $%
151.9289$ &  & $1309.2366$ &  & $1310.1172$ &  &  \\ \hline\hline
\end{tabular}

$^{a}$PW results, $^{b}$LYP results, and $^{c}$\cite{Martins97}.%
%TCIMACRO{\TeXButton{E}{\end{table}}}%
%BeginExpansion
\end{table}%
%EndExpansion
\endgroup

\bigskip

\begingroup\squeezetable

%TCIMACRO{\TeXButton{B}{\begin{table}[htbp] \centering}}%
%BeginExpansion
\begin{table}[htbp] \centering%
%EndExpansion
%TCIMACRO{%
%\TeXButton{T-9}{\caption{Total energies ($E$) and excitation energies ($\Delta E$) of inner-shell
%excited states $1s^{-1}ns$ $^{1,3}S$ ($n=4\thicksim 8$) of Mg.
%\label{T-9}}}}%
%BeginExpansion
\caption{Total energies ($E$) and excitation energies ($\Delta E$) of inner-shell
excited states $1s^{-1}ns$ $^{1,3}S$ ($n=4\thicksim 8$) of Mg.
\label{T-9}}%
%EndExpansion
\begin{tabular}{ccccccccccc}
\hline\hline
&  & $-E$ &  & (a.u.) &  & $\Delta E$ &  &  &  & (eV) \\ 
\cline{3-5}\cline{7-11}
States &  & Present &  & work &  & Present &  & work &  & Other theory \\ 
\cline{3-5}\cline{7-9}\cline{11-11}
&  & PW$^{a}$ &  & LYP$^{b}$ &  & PW$^{a}$ &  & LYP$^{b}$ &  & CIIOO$^{c}$
\\ \hline
\multicolumn{1}{r}{$1s^{-1}4s$ $^{3}S$} &  & $151.9980$ &  & $152.0171$ &  & 
$1306.7666$ &  & $1307.7160$ &  & 1306.29 \\ 
\multicolumn{1}{r}{$\ \ \ \ ^{1}S$} &  & $151.9957$ &  & $152.0144$ &  & $%
1306.8281$ &  & $1307.7895$ &  &  \\ 
\multicolumn{1}{r}{$1s^{-1}5s$ $^{3}S$} &  & $151.9442$ &  & $151.9649$ &  & 
$1308.2314$ &  & $1309.1362$ &  & 1308.77 \\ 
\multicolumn{1}{r}{$\ \ \ \ \ \ \ \ \ \ ^{1}S$} &  & $151.9434$ &  & $%
151.9640$ &  & $1308.2521$ &  & $1309.1618$ &  &  \\ 
\multicolumn{1}{r}{$1s^{-1}6s$ $^{3}S$} &  & $151.9243$ &  & $151.9454$ &  & 
$1308.7716$ &  & $1309.6660$ &  & 1309.35 \\ 
\multicolumn{1}{r}{$\ \ \ \ \ \ \ \ \ \ ^{1}S$} &  & $151.9240$ &  & $%
151.9450$ &  & $1308.7808$ &  & $1309.6780$ &  &  \\ 
\multicolumn{1}{r}{$1s^{-1}7s$ $^{3}S$} &  & $151.9146$ &  & $151.9360$ &  & 
$1309.0366$ &  & $1309.9226$ &  &  \\ 
\multicolumn{1}{r}{$\ \ \ \ \ \ \ \ ^{1}S$} &  & $151.9144$ &  & $151.9358$
&  & $1309.0421$ &  & $1309.9297$ &  &  \\ 
\multicolumn{1}{r}{$1s^{-1}8s$ $^{3}S$} &  & $151.9091$ &  & $151.9307$ &  & 
$1309.1868$ &  & $1310.0663$ &  &  \\ 
\multicolumn{1}{r}{$\ \ \ \ \ \ \ ^{1}S$} &  & $151.9089$ &  & $151.9306$ & 
& $1309.1901$ &  & $1310.0706$ &  &  \\ \hline\hline
\end{tabular}

$^{a}$PW results, $^{b}$LYP results, and $^{c}$\cite{Martins97}.%
%TCIMACRO{\TeXButton{E}{\end{table}}}%
%BeginExpansion
\end{table}%
%EndExpansion
\endgroup

\section{Conclusions}

In summary, the procedure we have developed for excited-state calculation
based on SLHF density functional approach and Slater's diagonal sum rule has
been extended to the treatment of inner-shell excited states of atomic
systems. In this procedure, electron spin-orbitals in an electronic
configuration are obtained first by solving the KS equation with the exact
SLHF exchange potential. Then a single-Slater-determinant energy of the
electronic configuration is calculated by using these electron
spin-orbitals. Finally, a multiplet energy of an excited state is evaluated
from the single-Slater-determinant energies of the electronic configurations
involved in terms of Slater's diagonal sum rule. In this procedure, the key
part is the SLHF exchange potential. This potential qualifies for
inner-shell excited-state calculation because it provides a potential with
free self-interaction, correct long-range behavior, and symmetry dependence
of atomic state. We have applied this procedure to the calculations of total
energies and excitation energies of inner-shell excited states of
close-shell atomic systems: Be, B$^{+}$, Ne, and Mg. In the calculation, the
generalized pseudospectral method with \textit{nonuniform} grids is used for
optimal discretization of the spatial coordinates, allowing accurate and
efficient treatment of the KS equation and the spin-orbital energies for
both the ground and excited states. The correlation effect is taken care of
by incorporating the PW and LYP correlation potentials and energy
functionals into calculation. The results from our calculations with LYP and
PW energy functionals are in overall good agreement with each other and also
with the available more sophisticated \textit{ab initio} theoretical results
and experimental data. The maximum relative discrepancy of our calculated
excitation energies to the available experimental results is not more than
0.90\%, demonstrating that the SLHF density-functional approach is capable
of providing a powerful and computationally efficient scheme for accurate
inner-shell excited-state calculation of close-shell atomic systems within
DFT. Extension of the SLHF density-functional approach to open-shell atomic
systems is in progress.

\begin{acknowledgments}
This work is partially supported by the Chemical Sciences, Geosciences and
Biosciences Division of the Office of Basic Energy Sciences, Office of
Science, U. S. Department of Energy, and by the National Science Foundation.
\end{acknowledgments}

\bigskip

\newpage

\bibliographystyle{apsrev}
\bibliography{dft2}

\end{document}